# Surface figure correction using differential deposition of WSi$_2$

Ch. Morawe, P. Bras, S. Labouré, F. Perrin, A. Vivo

ESRF, Grenoble, France

**ABSTRACT**

The surface figure of an x-ray mirror was improved by differential deposition of WSi$_2$ layers. DC magnetron sputtering through beam-defining apertures was applied on moving substrates to generate thin films with arbitrary longitudinal thickness variations. The required velocity profiles were calculated using a deconvolution algorithm. Height errors were evaluated after each correction iteration using off-line visible light surface metrology. WSi$_2$ was selected as a promising material since it conserves the initial substrate surface roughness and limits the film stress to acceptable levels. On a 300 mm long flat Si mirror the shape error was reduced to less than 0.2 nm RMS.

## 1. INTRODUCTION

The performance of reflective X-ray optics depends on the mirror quality. Surface figure errors alter the shape of the reflected wave front, which causes parasitic contrast in the flat field or blurring of the focal spot. In modern hard X-ray light sources such as the ESRF [1], figure accuracies down to 1 nm are required to preserve the source properties [2]. To correct for figure errors precise metrology and deterministic polishing techniques such as elastic emission machining (EEM) [3], ion beam figuring (IBF) [4,5], and differential deposition have been developed. EEM and IBF rely on material removal, differential deposition adds material by thin film coating. It has been developed within the astronomy community [6], but has also been applied to synchrotron optics [7-11]. This work deals with the correction of long to short period height errors of X-ray mirrors with a typical length of about 300 mm. The method and the experimental techniques will be described and the results of a differentially coated mirror will be shown. The potential impact of roughness and film stress on the corrected substrates will be discussed and directions for further development will be indicated.

## 2. METHOD

### 2.1 Theory

The differential deposition technique is based on a substrate that moves in front of a particle source following a specific velocity profile $v(x_m)$. The resulting film thickness distribution $t(x_s)$ can be expressed as [12]

$$t(x_s) = \int_{-S}^{+S} R \cdot f(x_m - x_s) \cdot \frac{dx_m}{v(x_m)} \tag{1}$$

where $f(x_m-x_s)$ is the normalized static flux profile of the source on the substrate and $R$ the growth rate at its centre. $x_s$ and $x_m$ are the substrate and motion coordinates, respectively. The integration is carried out over the full length $2S$ of the substrate motion. In the present case, the thickness distribution $t$ is given and the speed profile $v$ needs to be calculated, which corresponds to a deconvolution process. Using discrete steps equation (1) can be written as a system of linear equations. An algorithm based on matrix inversion was developed to solve the problem. It includes sub-routines that were originally developed for astronomical image deconvolution by NASA [13]. The growth rate $R$ and the flux profile $f$ are experimental input parameters for the programme.

## 2.2 Thin film deposition

The coatings were made at the ESRF multilayer deposition facility [14] using DC magnetron sputtering. The deposition process took place in an Ar atmosphere at a working pressure of 0.1 Pa. $WSi_2$ was deposited from a compound target with a power of 200 W. The growth rate was about 0.32 nm/s. Two beam defining apertures with openings of 2 mm and 24 mm were placed about 3 mm in front of the substrate surface and about 89 mm away from the $WSi_2$ sputter target. Stationary $WSi_2$ coatings, where the thickness is controlled by opening and closing a shutter, were carried out to measure the particle flux distribution in the substrate plane. The corrective coatings were made in dynamic mode where the substrate moves in front of the aperture following a pre-programmed speed profile. Repeated duty cycles were applied to obtain the required thickness profile. All test samples used for roughness and stress investigations were coated on thin Si wafers.

## 2.3 X-ray reflectivity (XRR)

The coatings were characterized on a laboratory X-ray reflectometer at 8048eV [15]. Specular reflectivity scans can be carried out with a dynamical range of up to $10^7$. To measure thin films with strongly varying thickness the X-ray beam can be oriented perpendicular to the thickness gradient and slit down to minimize the averaging impact of the beam footprint on the reflectivity data. Simulation software based on the Parratt formalism [16] allows for the precise determination of thicknesses, mass densities, and interface widths.

## 2.4 Surface metrology

The surface figure was measured with the ESRF Long Trace Profiler (LTP) [17], an in-house pencil beam deflectometer used to perform meridional measurements on up to 1.4 meter long surfaces, with an accuracy of 0.1 µrad and a lateral resolution of 2 mm. The mirror coordinate system reference is obtained with a precision better than 50 µm using the LTP signal detection. The mirror is supported by two rods spaced by half of its length in order to subtract gravity effects. The integration of the slope profile allows to retrieve the height profile of the mirror. In the present case, the residual shape error profile is obtained by subtracting a second order polynomial from the measured height profile.

## 2.5 In-situ stress measurements

The mechanical stress of the thin films was studied by recording the change of the macroscopic sample curvature and by applying the Stoney equation [18]. The curvature evolution was measured quasi in-situ using a specific monitor [19] and by coating a sequence of thin sub-layers of each concerned material.

## 3. RESULTS

### 3.1 Surface roughness

Initial studies [20] have shown that the use of single Cr films for corrective coatings is limited to relatively thin layers of about 20 nm due to rapidly increasing surface roughness. As an alternative, the properties of Pt and [C/Pt] multilayers (MLs) were explored [21]. Although these materials perform better than Cr, in particular [C/Pt] MLs, further improvements would be desirable. Fig.1a shows the root mean square (RMS) roughness evolution of Cr (red), Pt (blue), $[C(5nm)/Pt(20nm)]_N$ (green), and $WSi_2$ (black) versus thickness, as derived from XRR scans. While the roughness of Cr, Pt, and [C/Pt] increases with thickness, though on different absolute scales, the surface roughness of $WSi_2$ stays at a low level and even slightly decreases with thickness up to $t(WSi_2) > 300$ nm.

To verify the impact of the initial surface roughness on the quality of ML over-coatings, $[W/B_4C]_{20}$ MLs with a d-spacing of 2.5 nm were deposited on top of a selection of the samples shown in Fig.1a. In Fig.1b the average ML interface roughness is plotted as a function of the buffer layer thickness corresponding to Fig.1a. The evolution is similar to what was observed for the underlying single films. Only those MLs coated on $WSi_2$ buffer layers preserve or even improve their interface quality with increasing thickness. The absolute values are below 0.3 nm RMS and clearly better than what was achieved with other buffer layer materials. In terms of roughness, $WSi_2$ appears to be a promising candidate for differential deposition.

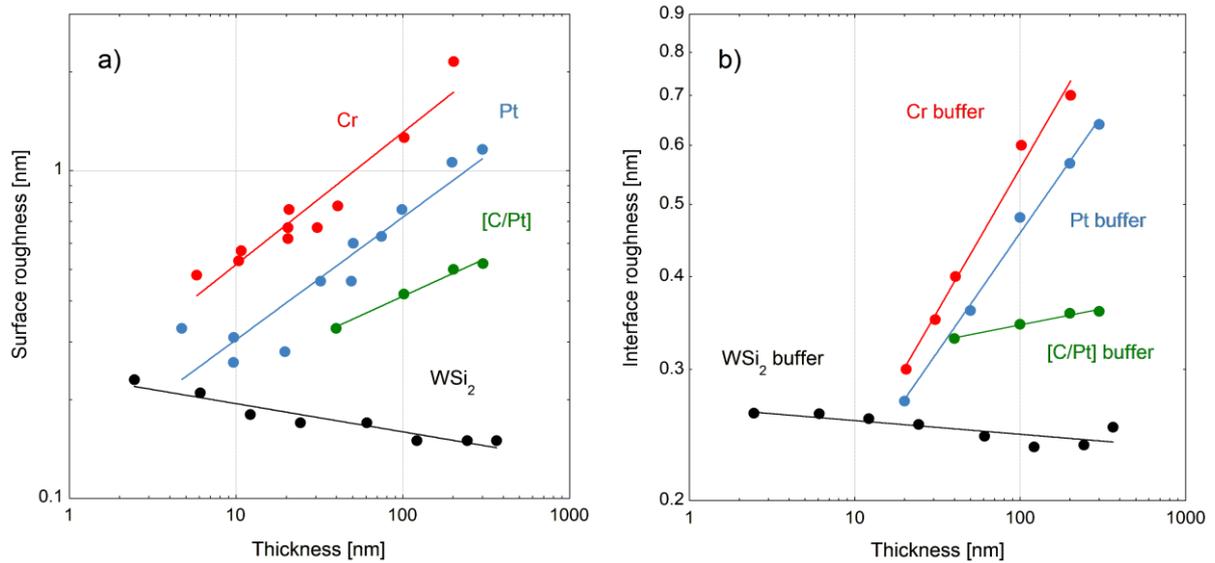

Fig.1: a) RMS surface roughness of Cr (red), Pt (blue), [C/Pt] (green), and WSi$_2$ (black) versus total film thickness. b) Average RMS interface roughness of [W/B$_4$C]$_{20}$ on Cr (red), Pt (blue), [C/Pt] (green), and WSi$_2$ (black) versus buffer layer thickness. Straight lines are fits using power laws.

### 3.2 Film stress

Elevated stress in thin films can cause substrate deformation or lead to cracks or delamination of the coatings. The integral stress of WSi$_2$/Si was measured in-situ versus thickness and compared with previous data of Cr/Si and Pt/Si as summarized in Fig.2. Cr grows tensile over the measured range up to t = 300 nm while Pt generates compressive stress, both with absolute stress levels of the order of $|\sigma|$ = 1.0 GPa. WSi$_2$ initially grows tensile but turns to compressive stress beyond t = 5 nm, then saturates near -0.5 GPa for t > 100 nm. Although this stress level remains considerable, the situations appears less critical than in the case of Cr or Pt. Therefore, the choice of WSi$_2$ for differential deposition can be confirmed.

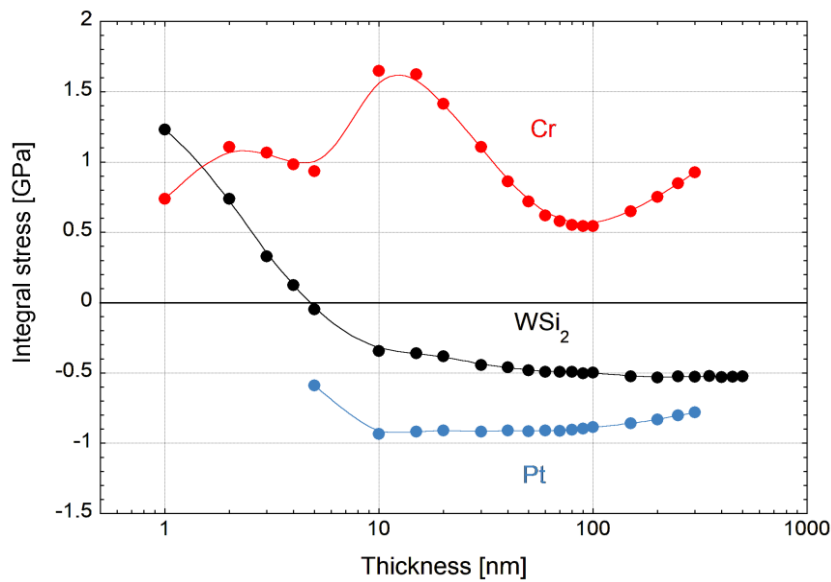

Fig.2: In-situ integral film stress of Cr/Si (red), Pt/Si (blue), and WSi$_2$/Si (black) versus thickness. Lines are guides to the eye.

In order to select optimum deposition conditions, both the surface roughness and the integral stress were measured as a function of the Ar pressure between 0.1 Pa and 1.0 Pa. The best results were obtained at low pressure, which is why all coatings were carried out at p(Ar) = 0.1 Pa.

### 3.3 Static flux distribution

To characterize the flux distribution two 20 nm thick $WSi_2$ films were deposited on stationary Si wafers through 2 mm and 24 mm wide apertures. Their local thickness was measured with XRR in 0.5 mm and 2 mm steps, respectively. The normalized thickness profiles are shown in Fig.3a for the 2 mm and in Fig.3b for the 24 mm opening. Both profiles show two pronounced shoulders that are edge projections of the two straight erosion lines on the rectangular sputter target.

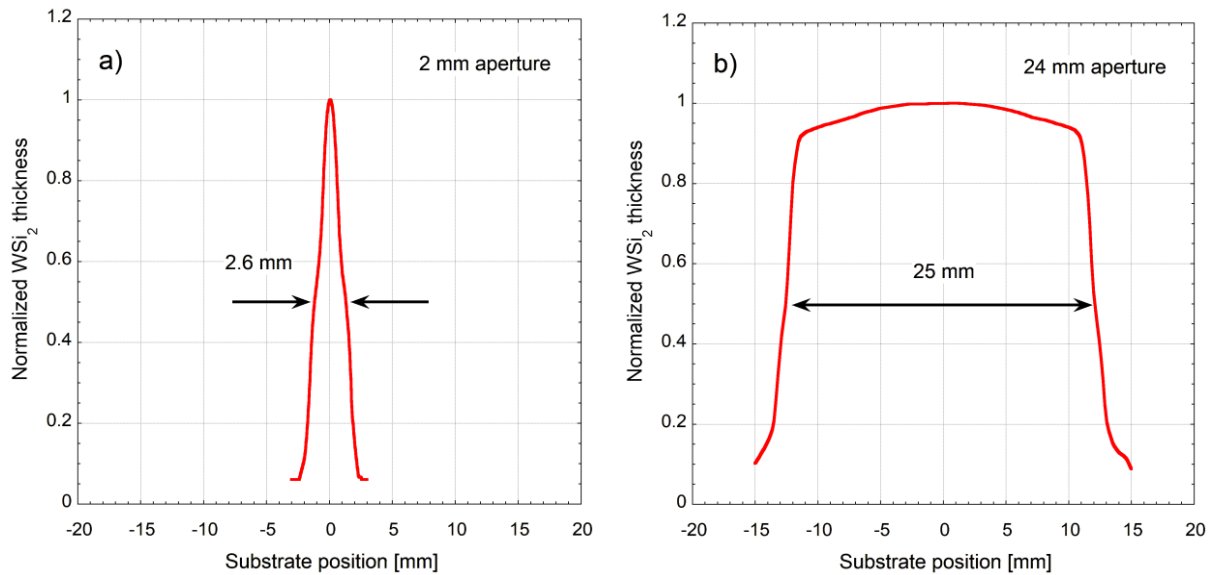

Fig.3: Normalized thickness profiles of $WSi_2$ films coated on stationary substrates through apertures of 2 mm (a) and 24 mm (b).

### 3.4 Figure correction of a Si mirror

A 300 mm long, 45 mm wide, and 30 mm thick Si mirror was selected for correction. Both the surface metrology and the differential deposition technique were applied to the central line along a trace of 280 mm. To provide identical surface conditions during all subsequent metrology studies, the Si substrate was first covered with a 30 nm thick uniform $WSi_2$ film and then measured with the LTP. The derived height error profile is displayed in Fig.4 (black curve). The initial height errors of this mirror were measured to be 3.87 nm (RMS) and 20.1 nm (PV).

A corrective coating with an average thickness of 20 nm was applied. During this process, the 24 mm aperture was inserted into the particle beam and only height errors extending over periods of more than 50 mm were attempted to be removed. The residual error profile after this first iteration is shown as the blue curve in Fig.4. As expected, long period variations have essentially been suppressed leaving only errors of shorter period. After this correction, the height errors were reduced to 0.57 nm (RMS) and 3.09 nm (PV).

A final correction with an average thickness of 4 nm was carried out, this time using the 2 mm aperture, with the aim to correct for errors over shorter periods. The outcome is given as the red curve in Fig.4. This time, the short period errors have been significantly attenuated. The residual height errors have dropped to 0.19 nm (RMS) and 1.58 nm (PV). After reduction of the analysis length from 280 mm to 260 mm, which is close to the useful aperture of the given mirror, the errors even drop to 0.14 nm (RMS) and 0.86 nm (PV).

All high frequency variations with periods below 5 mm remain visible in all three LTP measurements, which underlines the quality of the instrument. As expected, these very short period errors are not corrected with the present setup.

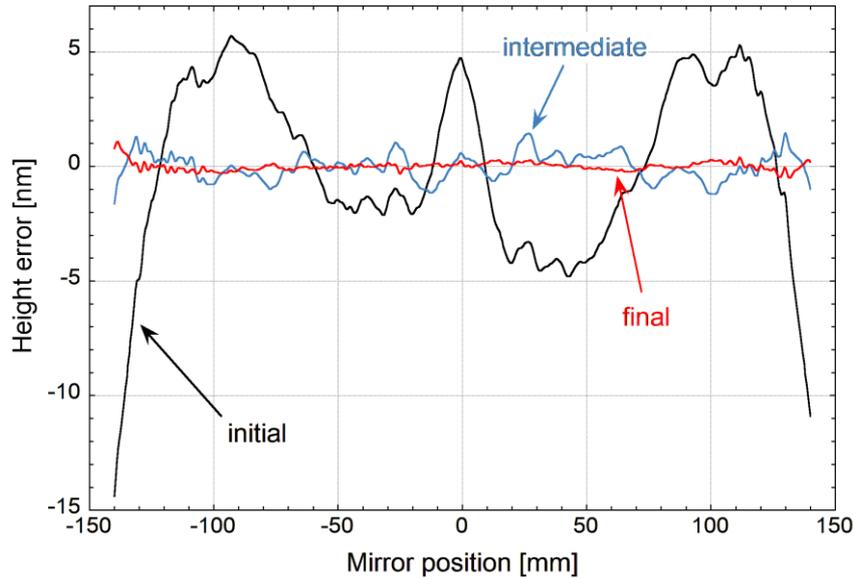

Fig.4: LTP measured surface height errors of the initial mirror (black), after intermediate long period (blue) and final short period (red) corrections.

## 4. DISCUSSION

### 4.1 Material properties

The choice to use $WSi_2$ films was motivated by its superior surface quality compared to layers of Cr, Pt, or [C/Pt]. The principal advantage of $WSi_2$ relies on the fact that it remains amorphous up to relatively high thicknesses while metals tend to crystallize beyond a critical thickness that depends on sputter parameters and growth conditions. $WSi_2$ was reported to maintain or improve the quality of an underlying Si surface [22]. $WSi_2$ generates moderate stress levels compared to Cr and Pt and it can be sputtered at relatively high rates from affordable compound targets. Therefore, $WSi_2$ appears to be the material of choice for differential deposition over a large thickness range.

### 4.2 Differential deposition and metrology

The evolution of the PV height errors along 280 mm after subsequent corrections is shown in Fig.5 on a logarithmic scale. The total improvement is more than 20 times. Only error periods below 5 mm can not be treated with the current experimental setup. The present positioning accuracy of the entire process, which is about ±0.3 mm, is sufficient for correction periods above 5 mm, but requires technical improvements when pushing to higher spatial frequencies. Some difficulties with parasitic visible light reflections from buried interfaces were observed, which perturb the LTP data. These issues will need to be addressed during the future development of this technique.

The new ESRF compact deposition system will provide improved positioning and motion accuracy. The use of narrower apertures would extend the correction capabilities to higher spatial frequencies. Combining differential deposition with subsequent ML coatings would enable tests on synchrotron beamlines in an operational environment.

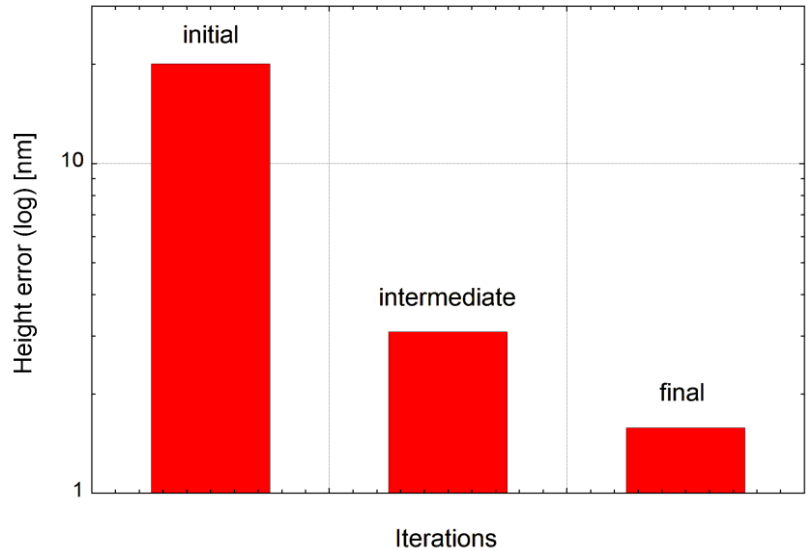

Fig.5: PV height errors on a logarithmic scale after subsequent iterations measured with the LTP over a length of 280 mm.

## 5. SUMMARY

A differential deposition technique based on magnetron sputtered $WSi_2$ films and off-line visible light metrology was developed to improve the figure of reflective X-ray optics on length scales between 5 mm and 300 mm. The choice of $WSi_2$ enables the deposition of films of low roughness and moderate intrinsic stress over a wide thickness range beyond 100 nm. A 24 mm aperture enables fast correction of shape errors with periods above 50 mm. With a 2 mm aperture, features down to 5 mm can be corrected. Along a 280 mm analysis length the surface figure of a Si mirror was improved by a factor of 20 down to 0.19 nm (RMS) and 1.58 nm (PV). Over a reduced length of 260 mm the figure errors drop to 0.14 nm (RMS) and 0.86 nm (PV), which is within the tight tolerances for X-ray mirrors to be deployed on 4[th] generation light sources.

## ACKNOWLEDGEMENTS

This project has received funding from the European Union´s Horizon 2020 research and innovation programme under grant agreement No. 101004728.